\documentclass{aa}

\title{Gravitational and magnetosonic waves in gamma-ray bursts.}
\author{J. Moortgat \and J. Kuijpers}
\institute{Department of Astrophysics, University of Nijmegen, PO Box 9010, 6500 GL Nijmegen, The Netherlands\\
\emph{e-mail:} moortgat, kuijpers@astro.kun.nl}
\offprints{J. Moortgat, \\\email{moortgat@astro.kun.nl}}
\date{Recieved 12 October 2002 /Accepted 30 January 2003}

\usepackage{amssymb, natbib}
\bibpunct{(}{)}{;}{a}{}{,}
\usepackage[english]{babel}
\usepackage{hyperref}
\usepackage{graphicx}

\abstract{
One of the possible sources of gamma-ray bursts ({\sc grb}s) are merging, compact neutronstar binaries. 
More than $90\%$ of the binding energy of such a binary is released in the form of gravitational waves ({\sc gw}s) in the last few seconds of the spiral-in phase before the formation of a black hole.
In this article we investigate whether a fraction of this {\sc gw} energy is transferred to magnetohydrodynamic waves in the magnetized plasma wind around the binary.
Using the $3$+$1$ orthonormal tedrad formalism, we study the propagation of a monochromatic, plane fronted, linearly polarized {\sc gw} perpendicular to the ambient magnetic field in an ultra-relativistic wind, first in the comoving and then in the observer frame. 
A closed set of general relativistic magnetohydrodynamic ({\sc grm}) equations is derived in the form of conservation laws for electric charge, matter energy, momentum and magnetic energy densities.
We linearize the {\sc grm} equations under the action of a monochromatic {\sc gw}, which acts as a driver and 
find that fast magneto-acoustic waves grow, with amplitudes proportional to the {\sc gw} amplitude and frequency and the strength of the background magnetic field. 
}

\begin{document}

\titlerunning{Gravitational and magnetosonic waves in {\sc grb}s.}
\authorrunning{J. Moortgat \& J. Kuijpers}		
\maketitle

\section{Introduction}
The coupling between gravitational waves and electromagnetic waves ({\sc emw}s) in a magnetized vacuum 
has been investigated extensively over the past $40$ years by a number of authors \citep{gertsenshtein, lupanov, fortini, zeldovich, gerlach}. These studies demonstrate the coherent excitation of {\sc emw}s by a monochromatic {\sc gw} propagating perpendicularly to a background magnetic field.

The first calculations including the presence of a plasma were done by \citet{macedo}, who found a coupling of {\sc gw}s to ordinary and extraordinary {\sc emw}s, whereas \citet{brodin99}  derived the parametric excitation of Langmuir waves by a {\sc gw} propagating through unmagnetized plasma.
In \citet{brodin00, brodinCQG}, the authors adopt the $3$+$1$ tedrad formalism \citep{membrane, ellis98} 
and show that the dispersion relation in a tenuous plasma differs only from the vacuum solution by a small wavenumber shift. In a plasma, however, subsequent non-linear conversions such as harmonic generation, might allow the {\sc emw} energy to escape as radiation with frequencies high enough to overcome the interstellar plasma frequency \citep{brodin0400}. 
A numerical estimate for the case of a merging {\sc ns}-{\sc ns} binary shows that the amplitude of the {\sc emw}s can be significant. 
Longitudinal waves are excited by higher-order {\sc gw}-{\sc emw} interactions \citep{brodin0600} and magnetosonic waves ({\sc msw}s) by {\sc gw}s propagating in a low-$\beta$ plasma \citep{papadopoulos2}.

In this article we consider {\sc ns}-{\sc ns} mergers as a source for {\sc grb}s and apply the $3$+$1$ formalism to the interaction of the {\sc gw}s emitted by the merger with the \emph{ultra-relativistic wind} of magnetized plasma around the binary. 
In the last seconds before the collapse to a black hole, a considerable fraction of $M_\odot c^2$ is released into this plasma in the form of {\sc gw}s \citep[see][ Table~1]{janka}.
 We show that these waves distort the extremely strong magnetic field, frozen into the plasma, and excite growing magnetosonic waves in the wind. Already before the merger, the binary is embedded in a relativistically expanding magnetized wind of (mainly) leptons from the orbiting neutronstars, so even a small transfer of {\sc gw} energy to the wind may provide an interesting \emph{central engine} mechanism to fuel a {\sc grb} fireball.

The outline of this article is as follows. In Sect. \ref{general}, the covariant expressions for the electromagnetic fields, the energy-momentum densities and the orthonormal tedrad for a linearly polarized {\sc gw} are recapitulated. A closed set of linearized {\sc grm} equations in the metric of a {\sc gw} is derived in Sect. \ref{mhd} and solved, first in the comoving frame of the plasma (Sect.  \ref{dispersion}) and then in the frame of an observer at rest with respect to the merger (Sect. \ref{observer}). A numerical example and the interpretation of our results are given in Sects. \ref{numerical} and \ref{discussion}, respectively. Sect. \ref{summary} comprises our conclusions.

Throughout Secs. \ref{general}--\ref{observer}, Gaussian geometrized units are adopted ($c=1$) and Latin indices stand for $0, 1, 2, 3$. In Sec. \ref{numerical}, however, the numerical results are given in {\sc si} units.

\section{Covariant fluid equations}\label{general}

\subsection{Electromagnetic field equations}\label{fieldequations}
Maxwell's equations in terms of the electromagnetic field tensors and the $4$-current density $j_\mathrm{m}^b = (\tau, \vec{j})$ are: 
\begin{equation}\label{maxwell}
\nabla_b F^{ab} = 4 \pi j_\mathrm{m}^a
\mbox{\hspace{0.5 cm}and\hspace{0.5 cm}} 
\nabla_b {\cal F}^{ab} =0,
\end{equation}
where the covariant Faraday tensor $F^{ab}$ and its dual ${\cal F}^{ab}$, can be decomposed into 4-vectors that, in the \emph{rest frame} of an observer with 4-velocity $u^a$, reduce to the electric and magnetic field strengths, $E^a = (0,\vec{E})$ and $B^a = (0,\vec{B})$ \citep{lichnerowicz}:
\begin{equation}\label{fab}
\begin{array}{lclclcl}
F^{ab} &=& u^a E^b\! -\! u^b E^a \!+ \!\epsilon^{abcd} B_c u_d&, &
{\cal F}^{ab} & \equiv& \frac{1}{2} \epsilon^{abcd} F_{cd}, \\[0.25cm]
E^a &\equiv  & F^{ab} u_b  &,&  B^a &\equiv & {\cal F}^{ab} u_b. 
\end{array}
\end{equation}
In ideal {\sc mhd}, the electric field vanishes in the rest frame of the plasma: $\vec{E} =0$. Therefore $E^a  \equiv F^{ab} u_b =0$ in any frame and Faraday's tensor reduces to: $F^{ab} =\epsilon^{abcd} B_c u_d $.

\subsection{Energy and momentum conservation}\label{energyconservation}
Energy and momentum conservation follow from the $4$-derivative of the energy-momentum tensor. For a magnetized ideal plasma this can be expressed in terms of the proper matter energy density $\mu$, the pressure $p$ and the metric $g^{ab}$, as \citep{weinberg, hawkingellis}:
\begin{eqnarray}\nonumber
T^{ab} &=& 
(\mu + p)u^{a} u^{b} + p g^{ab}  
+ \frac{1}{4\pi} (F^{a}_{\ c} F^{bc}\! -\! \frac{1}{4} g^{ab}  F^{cd}F_{cd}) \\\label{tmn}
\nabla_b T^{ab} &=& 
\nabla_b \left[(\mu + p)u^{a} u^{b} + p g^{ab}\right] - F^{ab} j_{\mathrm{m}b} = 0, 
\end{eqnarray}
where the current density satisfies Eq.~(\ref{maxwell}).

\subsection{Tedrad system for a gravitational wave}\label{tedrad}
In the linearized theory of gravity one usually splits the metric: $g^{ab} = \eta^{ab} + h^{ab}$, where $\eta^{ab} = \mbox{diag}(-1, 1, 1, 1)$ is the Minkowski metric, and $h^{ab}$ is the space-time perturbation caused by the {\sc gw}. 
For a transverse-traceless, linearly ($+$) polarized, monochromatic {\sc gw} with frequency $\omega_\mathrm{g} = k_\mathrm{g}$ propagating in the $z$-direction, this field satisfies: $h^{ab} = \mbox{diag}(0, h, -h, 0)$, with $h(z\!-\!t) = h \mathrm{e}^{i \omega_\mathrm{g} (z-t)}$ \citep{gravitation}.
 
The {\em proper} (observer) {\em reference frame}, however, comoving with freely moving bodies is defined with respect to the natural orthonormal tedrad \citep{brodin00}:
\begin{eqnarray}\nonumber
\begin{array}{lclclcl}
 e_{(0)}\!^a &= & (\partial_t, 0, 0, 0) &,& e_{(1)}\!^a &=& (0, (1-\frac{h}{2})\partial_x, 0, 0),\\ 
e_{(2)}\!^a &= & (0, 0, (1+\frac{h}{2})\partial_y, 0) &,& e_{(3)}\!^a &=& (0, 0, 0, \partial_z).
\end{array}
\end{eqnarray}
Decomposed with respect to this tedrad, the metric reduces to that of flat space: $g^{(ab)} = \eta^{ab}$ and the {\sc grm} equations closely resemble their Newtonian equivalents.

\section{Magnetohydrodynamics in the comoving frame}\label{mhd}
The physical situation we want to consider is that of a perfectly conducting, ideal plasma in the presence of a background magnetic field along the $x$-axis, perpendicular to the direction of {\sc gw} propagation (Fig.~\ref{nacfig}). First we study the plasma rest-frame where:
$\vec{B}^{(0)} = B_0 \vec{e}_x$, $\mu^{(0)} = \rho$ and $\vec{v}^{(0)}\!=\!\tau^{(0)}\!=\!p^{(0)}\!=\!\vec{j}_\mathrm{m}^{(0)}\!=\!\vec{E}^{(0)}\!=\!0$.

The effect of the {\sc gw} is to induce small perturbations in all these quantities. Therefore, all equations will be linearized around the unperturbed state.

\subsection{Maxwell's equations}
The relevant, linearized Maxwell equations in the specified tedrad are \citep{brodin00}:
\begin{eqnarray}\label{ampere}
\nabla \times \vec{B}^{(1)}  - \frac{\partial \vec{E}}{\partial t}^{(1)} & = &   4\pi \vec{j}^{(1)}_\mathrm{m} + \vec{j}^{(1)}_{E},  \\\label{faraday}
\nabla \times \vec{E}^{(1)} +  \frac{\partial \vec{B}}{\partial t}^{(1)} & = &  -\vec{j}^{(1)}_{B},
\end{eqnarray}
where the \emph{gravitationally induced current densities} are just the collected Ricci rotation coefficients or {\sc gw}-terms:
\begin{equation}\label{gwterms}
\vec{j}^{(1)}_\mathrm{E} \equiv
- \frac{B_0}{2} \frac{\partial h}{\partial z}^{(1)}\vec{e}_y
\mbox{\hspace{0.25 cm} and\hspace{0.25 cm}}
\vec{j}^{(1)}_\mathrm{B} \equiv - \frac{B_0}{2} \frac{\partial h}{\partial t}^{(1)}\vec{e}_x.
\end{equation}
The electric field can be eliminated by assuming the ideal {\sc mhd} approximation of a collisionless plasma (zero resistivity), where the electric field vanishes in the comoving frame and Ohm's law reduces to:
\begin{equation}\label{restohm}
\vec{E}^{(1)} = -\vec{v}^{(1)}\times\vec{B}^{(0)}.
\end{equation}

\subsection{Conservation equations}\label{conservation}
\emph{Charge continuity} follows readily from the antisymmetry of $F^{ab}$ and Eq.~(\ref{maxwell}): 
$\nabla_a(\nabla_b F^{ab}) = 4\pi \nabla_a {j}_\mathrm{m}^a = 0$. 

The evolution of the \emph{magnetic energy} $W$ and 
the \emph{Poynting flux} $\vec{S}$ are just projections of  Eqs.~(\ref{ampere}--\ref{faraday}) onto $\vec{B}^{(0)}$:
\begin{eqnarray}\label{magenergy}
\frac{\partial W}{\partial t}^{(1)} + \nabla\cdot \vec{S}^{(1)} &=& - W^{(0)}\frac{\partial h}{\partial t}^{(1)}, \\\label{poynting}
\frac{\partial \vec{S}}{\partial t}^{(1)} + \nabla W^{(1)} &=& W^{(0)} \frac{\partial h}{\partial z}^{(1)}\vec{e}_z  - \vec{F}^{(1)}_\mathrm{L},
\end{eqnarray}
with $\vec{F}_\mathrm{L}= \vec{j}\times\vec{B}$ the Lorentz force, $W^{(0)} = B_0^2/8\pi$ and: 
$$\begin{array}{lcl}
{W}^{(1)} = \frac{\vec{B}^{(0)}\cdot\vec{B}^{(1)}}{4\pi}, &
\vec{S}^{(1)}  = \frac{\vec{E}^{(1)}\times\vec{B}^{(0)}}{4\pi}, &
\vec{F}^{(1)}_{\mathrm{L}} = \vec{j}^{(1)}_\mathrm{m}\times \vec{B}^{(0)}.
\end{array}
$$
Eq.~(\ref{tmn}) results in \emph{particle} and \emph{momentum} conservation in terms of the momentum,  $\mbox{\boldmath $\pi$}^{(1)} = \mu^{(0)} \vec{v}^{(1)}$ and the spatial stress tensor $\tens{T}^{(1)}$:
\begin{eqnarray}\label{consmom}
\frac{\partial \mu}{\partial t}^{(1)} + \nabla\cdot\mbox{\boldmath $\pi$}^{(1)}= 0 &,&
\frac{\partial \mbox{\boldmath $\pi$}}{\partial t}^{(1)} + \nabla \cdot \tens{T}^{(1)}= 0, 
\end{eqnarray}
or, equivalently, as a linearized equation of motion ({\sc eom}):
\begin{equation}\label{eom}
\mu^{(0)} \frac{\partial \vec{v}}{\partial t}^{(1)} + c_s^2 \nabla{\mu}^{(1)}= \vec{F}^{(1)}_\mathrm{L},
\end{equation}
in terms of the sound velocity $c_s^2 \equiv \Gamma ({p}^{(1)}/{\mu}^{(1)})$, where $\Gamma $ is the polytropic index.
To first order, the {\sc eom} does not contain any {\sc gw} terms and the coupling of the {\sc gw} to the plasma occurs only through the Lorentz force as can be seen from Eqs.~(\ref{poynting}, \ref{eom}) (or equivalently, through the current density which couples the {\sc eom} to Maxwell's equations (\ref{ampere}--\ref{faraday})).

\section{Wave solutions in the comoving frame}\label{dispersion}

\subsection{Laplace transforms}
The proper way to handle the unstable response of a plasma to a disturbance is to make use of \emph{Laplace transforms} \citep{landau, melrose}. By giving the frequency or wavenumber a small positive imaginary part, one allows for damping or  growth and a way to deal with singularities due to resonance.
In an initial value problem it is customary to Laplace transform with respect to time, whereas in a boundary value problem, as we are considering here, a Laplace transform with respect to space is more suitable. These are defined by:  
\begin{equation}\label{lpctrnasform}
\begin{array}{lcl}
F(s) \equiv \int_0^\infty \!  \mathrm{e}^{- s z} f(z) \mathrm{d}z&,&
f(z) \equiv \int_{-i\infty + \eta}^{i\infty + \eta}\!  \mathrm{e}^{s z} F(s)\mathrm{d}s,
\end{array}
\end{equation}
where $\eta $ is an arbitrary positive constant chosen so that the contour of integration lies to the right side of all singularities in $F(s)$.
Below, we will transform with respect to $s \equiv i k$, where $\Re [k] \approx k_g = \omega_g$, so the contour of integration follows the real axis except for the poles at $k = \pm \omega_g /u_\mathrm{A}$ and $k = \omega_g$, where it deviates infinitesimally into the \emph{upper} half imaginary plane. 

For the time dependence we use Fourier transforms, which implies that the perturbations oscillate at the frequency of the driving {\sc gw}: $\propto \exp{(-i\omega_\mathrm{g} t)}$.

\subsection{Dispersion relations and wave solutions}\label{disper}
We will assume that the tenuous, strongly magnetized leptoid surrounding the binary has a low plasma-beta 
$\beta_\mathrm{pl} \!=\!(c_s/v_\mathrm{A})^2\!=\!4\pi p /B_0^2\! \ll \!1$, where the gas pressure is negligible with respect to the magnetic pressure. The classical Alfv\'en speed $v_\mathrm{A}^2 \equiv B_0^2/(4\pi\rho) \gg1$ so the displacement current is important. Perturbations propagate in the relativistic plasma with the \emph{generalized Alfv\'en speed}, $1/u_\mathrm{A}^2 \equiv 1 + 1/v_\mathrm{A}^2$. 

In Laplace and Fourier space the $13$ {\sc grm} equations derived in Sect. \ref{conservation} reduce to an algebraic system with $5$ non-trivial solutions:\footnote{Since the only zeroth order quantities are $B_0$ and $\rho $, the superscripts indicating the perturbations will be omitted.}
\begin{equation}\label{somsols}
\begin{array}{lcl}
v_z (\omega, k) &=& \frac{\omega}{k} \frac{\mu(\omega, k)}{\rho}  = \frac{B_0 j_y(\omega, k)}{i\omega \rho}  = 
\frac{(\omega + k)u_\mathrm{A}^2}{\omega + k u_\mathrm{A}^2} \frac{B_x (\omega, k)}{B_0} = \\[0.25cm]
-\frac{E_y (\omega, k) }{B_0} &=& \frac{i h \omega}{2} \frac{u_\mathrm{A}^2}{\omega^2 -k^2 u_\mathrm{A}^2}\frac{\omega + k}{\omega -k} \delta (\omega -\omega_\mathrm{g}).
\end{array}
\end{equation}
The inverse transformations lead to:
\begin{equation}\label{comovingsols}\begin{array}{lcl}
\frac{E_y(z,t)}{u_\mathrm{A} B_0} &=&  \\[.25cm]
- \frac{v_z(z,t)}{u_\mathrm{A}}&=& 
	\zeta \Re\left[\mathrm{e}^{i k (z-u_\mathrm{A} t)}
	\left\{1\! -\! \zeta_1 \mathrm{e}^{-i\Delta k z} \! -\!  
	\zeta_2 \mathrm{e}^{-2ik z}\right\}\right]\\[.25cm]
\frac{B_0 j_y(z,t)}{u_\mathrm{A}\rho \omega_g} &=& \zeta \Im\left[\mathrm{e}^{i k (z-u_\mathrm{A} t)}
	\left\{1\! -\! \zeta_1 \mathrm{e}^{-i\Delta k z} \! -\!  
	\zeta_2 \mathrm{e}^{-2ik z}\right\}\right] \\[.25cm]
-\frac{B_x(z,t)}{B_0} &=& \zeta\Re\left[\mathrm{e}^{i k (z-u_\mathrm{A} t)}
	\left\{1\! -\! \zeta_3 \mathrm{e}^{-i\Delta k z} \! +\!  
	\zeta_2 \mathrm{e}^{-2ik z}\right\}\right]\\[.25cm]
-\frac{\mu (z,t)}{\rho} &=& \zeta\Re\left[\mathrm{e}^{i k (z-u_\mathrm{A} t)}
	\left\{1\! -\! \zeta_4 \mathrm{e}^{-i\Delta k z} \! +\!  
	\zeta_2 \mathrm{e}^{-2ik z}\right\}\right]
	\end{array}\end{equation}
with $k u_\mathrm{A} = \omega_\mathrm{g}$, $\Delta k= k\!-\!\omega_\mathrm{g}$, $\xi \equiv u_\mathrm{A}/(1\!+\!u_\mathrm{A})^2$ and:
\begin{equation}\label{consts}
\begin{array}{lclcl}
\zeta = \frac{h}{4} \zeta_2^{-\frac{1}{2}} &,&
\zeta_1 =  4\xi &\mbox{\hspace{0.25 cm} $\gg$\hspace{0.25 cm}}&
\zeta_2 =u_\mathrm{A} \xi\left(\frac{\Delta k}{\omega_\mathrm{g}}\right)^2,\\[0.2cm]
\zeta_3 = 4u_\mathrm{A} \xi &,&
 \zeta_4 =  2\xi\frac{1\!+\!u_\mathrm{A}^2}{u_\mathrm{A}}
&\mbox{\hspace{0.25 cm} $\gg$\hspace{0.25 cm}}&
\zeta_2.
\end{array}
\end{equation}
Since $\Delta k$ is very small (see Sect. \ref{alfven}), we can expand the solutions around $\Delta k\! =\!0$, or equivalently $u_\mathrm{A} \!=\!1$, to find the dominant terms:
\begin{eqnarray}\label{restgrowth}\nonumber
\frac{B_x(z,t)}{B_0} &=& \frac{v_z(z,t)}{u_\mathrm{A}} = \frac{\mu (z,t)}{\rho}  =-\frac{E_y(z,t)}{u_\mathrm{A} B_0} \\
	&\simeq & \frac{h}{2} k z\  \Im\left[\mathrm{e}^{i k (z- u_\mathrm{A} t)} \right],\\\nonumber
\frac{B_0 j_y(z,t)}{u_\mathrm{A}\rho \omega_g} &\simeq& \frac{h}{2} k z\  \Re\left[\mathrm{e}^{i k (z- u_\mathrm{A} t)} \right].
\end{eqnarray}
These are \emph{fast magneto-acoustic waves}, propagating at the Alfv\'en speed in the same direction as the {\sc gw} and growing linearly with distance.

\section{Observer frame}\label{observer}

For a plasma flowing out relativistically in the $z$-direction with Lorentz factor 
$\gamma_\mathrm{tot} =1/\sqrt{1\!-\!(\beta\!+\!v_z)^2)} \simeq \gamma + \gamma^3\beta v_z $ ($\gamma \equiv \gamma_\beta$) corresponding to a constant velocity $\beta \approx 1$, the full set of linearized {\sc grm} equations is:
\begin{equation}
\label{relohm}
\begin{array}{lclr}
\left\{\frac{\partial }{\partial t} + \beta\frac{\partial }{\partial z}\right\}\mu &=&-  \gamma^2 \rho \left\{\beta \frac{\partial }{\partial t} +\frac{\partial }{\partial z}\right\}v_z
&\mbox{\scriptsize\sc continuity}\\[0.25cm]
\gamma^3 \rho \left\{\frac{\partial }{\partial t} + \beta \frac{\partial }{\partial z}\right\} v_z &=& -j_y B_0
&\mbox{\scriptsize\sc eom}\\[0.25cm]
\frac{\partial E_y}{\partial t} - \frac{\partial B_z}{\partial z} + 4\pi j_y&=&  \frac{i\omega_\mathrm{g} h}{2}  (B_0\! +\! E_0) \mathrm{e}^{i\omega_\mathrm{g} (z-t)}
&\mbox{\scriptsize\sc ampere}\\[0.25cm]
\frac{\partial E_y}{\partial z}  -\frac{\partial B_z}{\partial t} &=& \frac{i\omega_\mathrm{g} h}{2}  (B_0 \!+ \!E_0)  \mathrm{e}^{i\omega_\mathrm{g} (z-t)}
&\mbox{\scriptsize\sc faraday}\\[0.25cm]
E_y + v_z B_0 + \beta B_x &=& 0 \mbox{\hspace{0.25 cm} ,\hspace{0.25 cm}} E_0 = -\beta B_0
&\mbox{\scriptsize\sc ohm}
\end{array}
\end{equation}
where all quantities are now defined in the \emph{observer frame} at rest with respect to the binary (primes are used to indicate comoving quantities only where there is the risk of confusion). Explicitly: $\rho$, $\beta$, $B_0$ and $E_0=E_y^{(0)}$ are the zeroth order quantities and $\mu$, $v_z$, $B_x$, $E_y$, $j_y$, $h$ are the perturbations. 
The most important difference with respect to the comoving frame is the factor $(1\!-\!\beta)B_0$ in the {\sc gw} terms, due to the background electric field seen by the observer.

The solutions to the equilibrium system without the {\sc gw}-terms, are again $\propto 
\mathrm{e}^{i \omega_g (z/u_\mathrm{A} -t)}$ with
$
u_\mathrm{A} = 
\frac{\beta + u^\prime_\mathrm{A}}{1+\beta u^\prime_\mathrm{A}}
$
the Lorentz boosted, generalized Alfv\'en speed. 
The full system can be solved along the same lines as in Sect. \ref{disper} by straightforward calculation using Laplace and Fourier transformations. The physical differences between the comoving and the observer frames are clear from Eqs.~(\ref{relohm}).

A different approach -- leading to the same result -- is to Lorentz transform Eqs.~(\ref{comovingsols}). Since Lorentz transformations are linear transformations and furthermore the phase of a plane wave is an invariant, the general solutions have the same form as Eqs.~(\ref{comovingsols}), but with different amplitudes. 

The {\sc msw} components excited by a {\sc gw} propagating through a relativistically flowing plasma are given in terms of observer quantities by:
\begin{equation}\label{relsols}
\begin{array}{rcl}
\frac{E_y(z,t)}{u_\mathrm{A} B_0} &=& \Lambda \Re\left[\mathrm{e}^{i k (z-u_\mathrm{A} t)}
	\left\{1\! -\! \Lambda_1 \mathrm{e}^{-i\Delta k z} \! -\!  
	\Lambda_2 \mathrm{e}^{-2i\kappa z}\right\}\right]\\[.25cm]
-\frac{B_x(z,t)}{B_0} &=& \Lambda\Re\left[\mathrm{e}^{i k (z-u_\mathrm{A} t)}
	\left\{1\! -\! \Lambda_3 \mathrm{e}^{-i\Delta k z} \! -\!  
	\Lambda_4 \mathrm{e}^{-2i\kappa z}\right\}\right]\\[.25cm]
-\frac{v_z(z, t)}{u_\mathrm{A}-\beta} &=& \Lambda \Re\left[\mathrm{e}^{i k (z-u_\mathrm{A} t)}
	\left\{1\! -\! \Lambda_5 \mathrm{e}^{-i\Delta k z} \! -\!  
	\Lambda_6 \mathrm{e}^{-2i\kappa z}\right\}\right]\\[.25cm]
\frac{B_0 j_y(z,t)}{u_\mathrm{A}\rho \omega_g} &=& \Lambda \Re\left[\mathrm{e}^{i k (z-u_\mathrm{A} t)}
	\left\{1\! -\! \Lambda_7 \mathrm{e}^{-i\Delta k z} \! -\!  
	\Lambda_8 \mathrm{e}^{-2i\kappa z}\right\}\right]
\\[.25cm]
-\frac{\mu (z,t)}{\rho} &=& \lambda\Re\left[\mathrm{e}^{i k (z-u_\mathrm{A} t)}
	\left\{1\! -\! \Lambda_{9} \mathrm{e}^{-i\Delta k z} \! -\!  
	\Lambda_{6} \mathrm{e}^{-2i\kappa z}\right\}\right]
\end{array}\end{equation}
where the constants $\Lambda$--$\Lambda_{9}, \lambda$ are defined in Appendix \ref{constants}. 
As in Sect. \ref{disper}, the dominant terms are found by expanding around $\Delta k=0$ (justified in Sect. \ref{alfven}):
\begin{equation}\label{relgrowtha}
\begin{array}{lcl}
\frac{B_x(z,t)}{B_0} = -\frac{E_y(z,t)}{u_\mathrm{A} B_0} = \frac{v_z(z, t)}{u_\mathrm{A}\!-\!\beta} &=&\frac{\mu(z,t)}{\rho} = \\[.25cm]
\frac{(1\!-\!\beta)h}{2} k z \  \Im \left[\mathrm{e}^{ik(z-u_\mathrm{A} t)}\right]
&\simeq& \frac{h}{4} \frac{k z}{\gamma^2}  \ \Im \left[\mathrm{e}^{ik(z-u_\mathrm{A} t)}\right],\\[.25cm]
\frac{B_0 j_y (z, t)}{\hat{\rho} \omega_\mathrm{g} u_\mathrm{A}} &\simeq & \frac{h}{4} \frac{k z}{\gamma^2}  \ \Re \left[\mathrm{e}^{ik(z-u_\mathrm{A} t)}\right].
\end{array}\end{equation}
where: $\hat{\rho} \equiv \frac{B_0^2}{4\pi}\frac{1-u_\mathrm{A}^2}{u_\mathrm{A}^2}$. Apparently, the interaction is less efficient when the plasma is escaping relativistically, than when it is at rest with respect to the source of {\sc gw}s.

As a final remark: Eq.~(\ref{relgrowtha}) is equivalent to the result of an initial value approach, where the Laplace transformations are performed with respect to $t$ instead of $z$. The amplitudes in Eq.~(\ref{relgrowtha}) become proportional to $\omega t$ instead of $k z$, but the characteristic timescale $T$ is related to the size of the interaction region $R$ by: $\omega T \sim \omega (R/u_\mathrm{A}) = k R$.

\section{Numerical estimates}\label{numerical}

\begin{table*}
$$
\begin{array}{|rcl|rcl|rcl|rcl|}\hline
\multicolumn{6}{|c|}{\mbox{Lab/Observer frame}} & \multicolumn{6}{|c|}{\mbox{Plasma/Comoving frame}} \\\hline\hline
&&&&&&&&&&& \\[-0.3cm]
\omega_\mathrm{g} &= & k_\mathrm{g} c & 
k_\mathrm{g} &\sim& 2 \cdot 10^{-5} \mathrm{m}^{-1} &
\omega_\mathrm{g}^\prime \approx \omega_\mathrm{g}/(2\gamma) &= &k_\mathrm{g}^\prime c  & 
k_\mathrm{g}^\prime &\sim & 10^{-7}  \mathrm{m}^{-1}   \\[0.10cm]
&&&&&&&&&&& \\[-0.3cm]
\omega_\mathrm{\sc msw} &= & \omega_\mathrm{g} = k_\mathrm{\sc msw} u_\mathrm{A} & 
\omega_\mathrm{g} &\sim& 2 \pi \cdot 10^3 \mathrm{rad}\ \mathrm{s}^{-1}&
\omega^\prime_\mathrm{\sc msw} =  \omega_\mathrm{g}^\prime &= &k^\prime_\mathrm{\sc msw} u^\prime_\mathrm{A} & 
\omega_\mathrm{g}^\prime &\sim&30\ \mathrm{rad}\ \mathrm{s}^{-1}  \\[0.1cm]\hline
&&&&&&&&&&& \\[-0.3cm]
\rho_\star &=&  
\frac{M m_\mathrm{e}}{e}\left(\frac{\Omega_\mathrm{b} B_\star}{2\pi c}\right) & 
\rho_\mathrm{lc} &\sim& 4 \cdot 10^{-10} \mathrm{kg}\  \mathrm{m}^{-3}&
\frac{v_\mathrm{A}^{\prime 2}}{c^2} =
\frac{B^{\prime 2}}{4\pi \rho^\prime c^2} &= &\frac{B_\mathrm{lc}^2}{4\pi \gamma \rho_\mathrm{lc} c^2}=&
&&  \\[0.10cm]
&&&&&&&&&&& \\[-0.3cm]
\rho_\mathrm{lc}&=&\rho_\star \left(\frac{B_\mathrm{lc}}{B_\star}\right)
& \Omega_\mathrm{e}&=&\frac{e B_\mathrm{lc}}{m_\mathrm{e} c}\sim 7 \cdot 10^{15} \mathrm{rad}\ \mathrm{s}^{-1}&
&=&\frac{\Omega_\mathrm{e}}{2\gamma_\mathrm{p} \Omega_\mathrm{b}} &
v_\mathrm{A}^\prime &\sim& 570 c\\[0.20cm]\hline 
&&&&&&&&&&& \\[-0.3cm]
u_\mathrm{A} &=& \frac{u_\mathrm{A}^\prime + \beta c}{1+\beta u_\mathrm{A}^\prime/c} &  
&& &
\frac{u_\mathrm{A}^{\prime 2}}{c^2} &=& 
\frac{v_\mathrm{A}^{\prime 2}}{c^2+v_\mathrm{A}^{\prime 2}}
& && \\[0.10cm]
&&&&&&&&&&& \\[-0.3cm]
\gamma_{u_\mathrm{A}} &\approx & 2 \gamma \gamma_{u_\mathrm{A}^\prime} &
\gamma_{u_\mathrm{A}} &\sim& 10^5 &
\gamma_{u_\mathrm{A}^\prime}^2 &\approx & \frac{\Omega_\mathrm{e}}{2 \gamma_\mathrm{p} \Omega_\mathrm{b}} &
\gamma_{u_\mathrm{A}^\prime} &\sim& 570 \\[0.2cm]\hline
&&&&&&&&&&& \\[-0.3cm]
 \Delta k &=& \frac{\omega_\mathrm{g}}{c} \left(\frac{c}{u_\mathrm{A}}\! -\!1\right)  & 
&&  &
\Delta k^\prime \equiv k^\prime\!-\! k_\mathrm{g}^\prime &=& \frac{\omega_\mathrm{g}^\prime}{c} \left(\frac{c}{u_\mathrm{A}^\prime}\!-\!1\right) &
&& \\[0.1cm]
&&&&&&&&&&& \\[-0.3cm]
&=&\frac{\Delta k^\prime}{2\gamma} & 
\Delta k &\sim& 8\cdot 10^{-16} \mathrm{m}^{-1} & 
&\approx & \frac{\omega_\mathrm{g} M}{2 c} \frac{\Omega_\mathrm{b}}{\Omega_\mathrm{e}} & 
\Delta k^\prime &\sim& 1.6\cdot 10^{-13} \mathrm{m}^{-1}\\[0.2cm]\hline
&&&&&&&&&&& \\[-0.2cm]
R \Delta k&\ll&1 &
R_\mathrm{max} &\sim& \frac{0.075}{\Delta k} \sim 10^{14}\mathrm{m} &
 && &&&\\[0.1cm]\hline\hline
\end{array}
$$
\caption{Formulas and numerical values in the frames at rest with respect to an observer and the plasma, respectively. In the numerical examples of this table we have taken: $B_\star = 10^9$T, $\gamma = \gamma_\mathrm{s} =100$, $M=10^5$, $\Omega_\mathrm{b}/2\pi = 160$Hz.}
\label{tabel}
\end{table*}

\noindent
In this section the results obtained in the previous sections are applied to the environs of a {\sc ns}-{\sc ns} binary close to merging. 
Both stars will have an electron-positron wind  filling the surrounding space with plasma up to large distances (whether the plasma is $e^\pm$ or baryon loaded is not important as long as it satisfies the ideal {\sc mhd} condition). The extent of the interaction region is determined by the distance $R_\mathrm{max}$ from the source where either the force-free or the $z \ll 1/\Delta k$ assumption brakes down. Within this scale height, we estimate the magnitude of the excited {\sc msw} amplitudes, including the decrease of the {\sc gw}-amplitude, the density and the magnetic field with distance.
A short numerical analysis is made for the magnetic fields, plasma densities, Lorentz factors and Alfv\'en velocities in the relativistic  plasma wind, summarized in Table \ref{tabel}. 
\paragraph{Note:} In this section proper dimensions in $c$ are restored and the numerical results are converted to {\sc si} units.

\subsection{Magnetic field configuration}
For pulsar-like neutronstars, the magnetic field close to the surface falls of as a dipole: $ B (r) = B_\star \left(R_\star/r\right)^3$.
In the {\sc mhd} approximation the plasma is `glued' to the field lines and forced to corotate up to the \emph{lightcylinder} where corotation requires superluminal velocities. Here, we consider a {\sc ns}-{\sc ns} merger where each {\sc ns} has its own magnetic field. The key element in the electromagnetic description of rotating magnetized stars is the deviation from inertial motion. In our case this is not the usual stellar rotation but the orbital motion combined with the individual stellar rotations. As the merger coalesces the orbital frequency increases and dominates over any other (rotational) motion.
Therefore, we assume that at the end of the spiral-in phase the \emph{orbital} rotation of the binary (with $\Omega_\mathrm{b} \sim 10^3$rad/s) determines the light-cylinder radius: $R_\mathrm{lc} = c/\Omega_\mathrm{b} \simeq 300$km. Here the field lines, anchored on the stellar polar caps, open up and the plasma is free to flow out along the field in a \emph{force-free wind} in which 
the toroidal component of the field dominates the poloidal component \citep{kuijpersequatorial}:

\begin{equation}\label{spiral}\begin{array}{lcl}\nonumber
B_\mathrm{t} (r\!>\!R_\mathrm{lc}) = B_\mathrm{lc} \left(\frac{R_\mathrm{lc}}{r}\right), \quad
B_{\mathrm{p}}(r\!>\!R_\mathrm{lc}) = B_\mathrm{lc} \left(\frac{R_\mathrm{lc}}{r}\right)^2, && \\[0.25cm]
B_\mathrm{lc} = B_\star (\frac{\Omega_\mathrm{b} R_\star}{c})^3 
\simeq  4 \cdot 10^{5} \mathrm{T} (B_{\star,9}) (\Omega_{\mathrm{b},3})^3 (R_{\star, 4})^3. &&
\end{array}\end{equation}

\subsection{Nature of the wind}
Above the polar caps of a pulsar the field lines are open, particles flow outwards and a steady charge density cannot be maintained at the \emph{Goldreich-Julian density} everywhere $n_\mathrm{GJ} = \vec{\Omega}_\mathrm{b}\!\cdot\!\vec{B}/(2\pi e c)$ \citep{beskin, lyubarsky}. 
As a result, a strong electric field develops along the magnetic field above the polar cap and charged `primary' particles are extracted from the surface with a density $n_\mathrm{p} \simeq n_\mathrm{GJ\star}$ and accelerated to high Lorentz factors (a typical number is $\gamma_\mathrm{p} \sim 10^7$). Note that the available potential jump is proportional to $\Delta \Psi \propto \Omega_\mathrm{b} B_\star $ in our case and can be much larger than for single pulsars.

Processes such as curvature radiation and inverse Compton emission then result in a cascade of  `secondary' $e^\pm$ pairs with a particle number density $n_\mathrm{s} = M n_\mathrm{p}$, where $M$ is called the \emph{multiplicity}. Due to energy conservation $n_\mathrm{p}\gamma_\mathrm{p} = n_\mathrm{s}\gamma_\mathrm{s}$, so the Lorentz factor of the secondary particles is $\gamma_\mathrm{s} = \gamma_\mathrm{p}/M \sim 100$ for $M \sim 10^5$. The secondary plasma flows out as a relativistic wind along the open magnetic field lines which have a dipole-like radial dependence up to  $R_\mathrm{lc}$, and develops into a spiral (Eq.~\ref{spiral}) further out. The charge density in the wind is adjusted to the local Goldreich-Julian density everywhere and the wind remains force-free up to a large distance.

At the lightcylinder the density is:
\begin{equation}\label{gj}\begin{array}{lcl}
 n_\mathrm{lc} $=$ n_\star \frac{B_\mathrm{lc}}{B_\star} = \frac{M \Omega_\mathrm{b} B_\mathrm{lc}}{2\pi e c} \sim 4\cdot 10^{20} \ \mathrm{m}^{-3}.
\end{array}\end{equation}

\subsection{Alfv\'en speeds \& Interaction lengthscale}\label{alfven}
The classical Alfv\'en speed $v_\mathrm{A}$ is proportional to $B_0/\sqrt{n}$. In the force free wind we have $B_0 \propto 1/r$ and $n \propto 1/r^2$, so $v_\mathrm{A}(r\!>\! R_\mathrm{lc})=$ constant. Therefore, we can evaluate the Alfv\'en velocities at the light-cylinder in the \emph{comoving frame} (using Eqs.~(\ref{gj})):
\begin{equation}\label{clasalfv}
\begin{array}{lcl}
B_\mathrm{lc}^\prime &=& \frac{B_\mathrm{lc}}{\gamma_\mathrm{s}}  \mbox{\hspace{0.5 cm}and\hspace{0.5 cm}} 
n_\mathrm{lc}^\prime = \frac{n_\mathrm{lc}}{\gamma_\mathrm{s}},  \\[.25cm]
\frac{v_\mathrm{A}^{\prime 2}}{c^2} &=&
 \frac{B_\mathrm{lc}^{\prime 2}}{4\pi n_\mathrm{lc}^\prime m_\mathrm{e} c^2} = 
 \frac{B_\mathrm{lc}^{2}}{4\pi \gamma_\mathrm{s} n_\mathrm{lc} m_\mathrm{e} c^2} 
= \frac{\Omega_\mathrm{e}}{2\gamma_\mathrm{p}\Omega_\mathrm{b}} \gg 1,  
\end{array}\end{equation}
in terms of the electron cyclotron frequency $\Omega_\mathrm{e} = \frac{e B_\mathrm{lc}}{m_\mathrm{e} c}$.
For the \emph{generalized} Alfv\'en speed and the wavenumber difference $\Delta k^{\prime}$, one finds (see also Table~\ref{tabel}):
\begin{equation}\label{generalalfv}
\begin{array}{lclclcl}
\frac{u_\mathrm{A}^{\prime 2}}{c^2} &=& \frac{v_\mathrm{A}^{\prime 2}}{c^2 + v_\mathrm{A}^{\prime 2}} \simeq 1 \mbox{\hspace{0.5 cm}or\hspace{0.5 cm}}
\gamma_{u_\mathrm{A}^\prime}^2 &\approx & \frac{\Omega_\mathrm{e}}{2 \gamma_\mathrm{p} \Omega_\mathrm{b}} \gg 1,
\\[.25cm]
\Delta k^{\prime} &\equiv&k^\prime - k_\mathrm{g}^\prime = \frac{\omega_\mathrm{g}^\prime}{c}\left(\frac{c}{u_\mathrm{A}^{\prime}}-1\right) &\ll& 1.
&
\end{array}\end{equation}
The equivalent quantities in the \emph{observer frame} are derived from Lorentz transformations and can be found in Table~\ref{tabel}. 
The important conclusion from these estimates is that the maximum distance over which the {\sc msw} amplitudes can grow is: $z \ll 1/\Delta  k \simeq 10^{15}\mathrm{m}$ (see Eq.~(\ref{relgrowtha})), which is comparable to the limits on the {\sc mhd} approximation as calculated by \citet{drenkhahn2001}.

\subsection{Magnitude of excited magnetosonic waves}
The frequency of {\sc gw}s emitted by a merging binary is related to the angular frequency of the last orbits by: $\omega_\mathrm{g}\! =\! 2 \Omega_\mathrm{b}\! \sim\! 4 \pi\cdot 10^3$~rad/s \citep{shibata}, so $R_\mathrm{lc}\! \sim\! 50$km and $B_\mathrm{lc}\!\sim \!10^7$T.    
We assume that for $R_\mathrm{in}\!\sim \!10 R_\mathrm{lc}$ we are in the far field of the merger and estimate the amplitude of $B^{(1)}_x (z,t)$ in Eq.~(\ref{relgrowtha}) at the start of the interaction region. Taking into account the (approximate) $1/r$ decrease of $h$ and $B_0$,
$$
\begin{array}{lcl}
h &\sim &10^{-3}\left(\frac{R_\mathrm{in}}{r}\right),\quad 
B_{0} \sim B_\mathrm{lc} \left(\frac{R_\mathrm{lc}}{R_\mathrm{in}}\right)\left(\frac{R_\mathrm{in}}{r}\right),
\end{array}
$$
we find:
\begin{equation}\label{torpol}
\begin{array}{lcl}
B^{(1)} (r\!>\!R_\mathrm{in}) &\sim & 0.5 \mathrm{T} \left(\frac{R_\mathrm{in}}{r}\right)\times [\frac{h_{-3}}{(\gamma_2)^2} k_{\mathrm{g},-5}B_{\mathrm{lc},7}R_{\mathrm{in}, 5}].
\end{array}
\end{equation}
The volume integrated energy of the {\sc msw} grows linearly with distance since $B^{(1)} \! \propto\! B_0 \propto\! 1/r$ and $V\!\propto \!r^3$. For an interaction region of $R_{\mathrm{max}}\!\sim\! 0.03$~pc we find for the magnetic component of the excited {\sc mhd} wave a total energy of:
\begin{equation}\label{energy}
\begin{array}{lcl}
 \tens{T}_\mathrm{B}^{(1)} &=& V\frac{B_0 B^{(1)}}{4\pi}  \sim 10^{37} \mathrm{J}, 
\end{array}
\end{equation}
which amounts to a fraction $10^{-6}$ of the magnetic energy $\tens{T}_\mathrm{B}^{(0)} = \frac{V B_0^2}{8\pi}  \sim 10^{43} \mathrm{J}$ of the Poynting flux dominated wind in the same volume.

\begin{figure}[h!]
\begin{center}
\resizebox{\hsize}{!}{\includegraphics{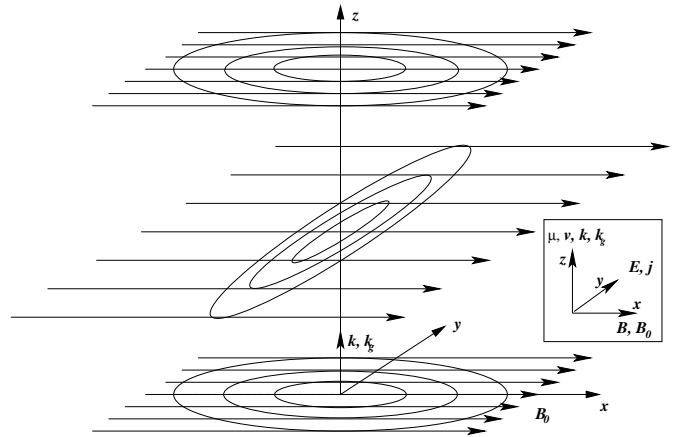}}
\caption{A {\sc gw} propagating in the positive $z$-direction across an ambient magnetic field (in the $x$-direction) excites a {\sc msw}. The orientations of the {\sc msw} components are indicated in the inset.}
\label{nacfig}
\end{center}
\end{figure} 

\section{Discussion}\label{discussion}
It is known that, to first order in $h$, {\sc gw}s propagating \emph{along} a uniform magnetic field do not couple to the field, neither in vacuum \citep{fortini}, nor in a plasma \citep{brodin0600, papadopoulos2} so we studied {\sc gw}-propagation \emph{perpendicular} to the magnetic field frozen into a plasma. 
An intuitive illustration of the process is given in Fig.~\ref{nacfig}, where a {\sc gw} traveling in the positive $z$-direction, deforms (imaginary) rings of plasma-particles in the $x$--$y$ plane into ellipses with long axes alternating periodically along the magnetic field direction ($x$-axis) and the $y$-axis. 
Consequently, the uniform field is periodically compressed and stretched leading to a \emph{modulated magnetic field strength} even though the {\sc gw} is divergence-free. 
The resulting magnetic pressure gradients in the $z$-direction try to re-establish a uniform field configuration and excite the compressional \emph{matter part} of the {\sc msw}. Since the plasma is glued to the field lines it is dragged along with the field. 
The velocity in the $z$-direction generates an \emph{electric field} ($-\vec{v}\times\vec{B}$) with a corresponding $\vec{E}\times\vec{B}$ \emph{drift-velocity}, whereas the magnetic gradient induces a $\vec{B}\times\nabla\vec{B}$ \emph{current density} and \emph{Lorentz force}.

The problem of {\sc gw} propagation in a plasma at rest was also studied by 
\citet{papadopoulos2} who found that the excitation of {\sc msw}s by a {\sc gw} is only possible when there is a wavenumber mismatch: $B_x^{(1)} \propto \Delta k$. This does not agree with our results (as presented in Sec. \ref{disper}), nor with those of \citet{brodin00} who show that the growth rate in a tenuous plasma smoothly matches that in a vacuum \citep{fortini}. 
Both from the {\sc grm} equations (Eqs. (\ref{magenergy}--\ref{eom}) or (\ref{relohm})) and from their solutions (Eqs. (\ref{restgrowth}) or (\ref{relgrowtha})) it is clear that in the vacuum limit $\mu^{(1)}\! \propto\! j_y^{(1)}\! \propto \!\rho\! \downarrow\! 0$ vanish as does $v_z^{(1)}$. The {\sc gw}s excite {\sc emw}s that propagate with the speed of light ($\Delta k\! \downarrow \!0$) and with amplitudes that can be obtained from Eq.~(\ref{restgrowth}) or Eq.~(\ref{relgrowtha}) by taking the limit $u_\mathrm{A}\! \rightarrow \!c$ and $k\! \rightarrow\! k_\mathrm{g}$ (and $\gamma=1$). This result is not surprising as we have taken into account both the material current and the vacuum displacement current.

\section{Conclusion}\label{summary}
The results presented in this paper are a first attempt to study the interaction of the relatively strong {\sc gw}s emitted by 
a {\sc grb} with an ultra-relativistic plasma wind. The space surrounding a merging {\sc ns}-{\sc ns} binary is already filled with such a wind up to large distances ($\sim 0.1$~pc). Moreover, in the merger almost all of the binding energy is released in the form of {\sc gw}s.

We derive a closed set of {\sc grm} equations both in the natural orthonormal measurement frame (the $3$+$1$ split) of the plasma and for an observer at rest with respect to the binary. These non-coordinate equations strongly resemble their Newtonian equivalents but have extra source terms due to the {\sc gw}. These gravity terms act as a driver for \emph{fast magnetosonic waves} with amplitudes that grow linearly with distance and are proportional to the {\sc gw}-frequency and amplitude and to the ambient magnetic field strength. 
It is the extended force-free wind in which the magnetic field only falls of as $1/r$ that provides the long interaction lengthscale. 

The total amount of energy that is transferred from the {\sc gw}s to the plasma, as given in Eq. (\ref{energy}), is substantial but for this case still much smaller than the average observed {\sc grb} energy. Note that for \emph{magnetars} one can have $B_\star \sim 10^{12}$~T and $\omega_\mathrm{g} \sim 15$~kHz, so that $(B_0 B^{(1)})/4\pi$ could be as much as a factor $10^7$ larger. However, for magnetars it is not obvious what to assume for the surrounding plasma.

In future work we will investigate what the observable effects of the {\sc gw}-{\sc msw} interaction are on the emitted radiation and its polarization and how a {\sc gw}-\emph{chirp} alters the results.

\begin{acknowledgements}
We would like to thank V.~S. Beskin, D. Papadopoulos, R. Pudritz, N. Stergioulas and L. Vlahos for useful discussions. This work was supported by the Dutch Research School for Astronomy, {\sc nova}.
\end{acknowledgements}

\appendix
\section{Constants in Eq.~(\ref{relsols})}\label{constants}
$$
\begin{array}{lcl}
\Lambda &=& \frac{h}{4} \frac{1+u_\mathrm{A}}{1- u_\mathrm{A}} \frac{(1-\beta)^2}{1-u_\mathrm{A}\beta} \simeq \frac{h}{4} \frac{\omega_\mathrm{g}}{\gamma^2\Delta k}  \\[0.25cm]
\Lambda_1 &=& 2\frac{1-u_\mathrm{A}\beta}{(1-\beta)^3}\frac{\beta(1-4u_\mathrm{A} + \beta) +u_\mathrm{A}^2(2-\beta(1-\beta))}{u_\mathrm{A}(1+u_\mathrm{A})^2} =1+ {\cal O}[\Delta k]^2\\[0.25cm]
\Lambda_2 &=& \gamma^2\left(\frac{1-u_\mathrm{A}}{1+u_\mathrm{A}}\frac{1+\beta}{1-\beta}\right)^2\frac{u_\mathrm{A}-2\beta +u_\mathrm{A}\beta^2}{u_\mathrm{A}} = {\cal O}[\Delta k]^2\\[0.25cm]
\Lambda_3 &=& 2\frac{1-u_\mathrm{A}\beta}{(1-\beta)^3} \frac{(1 + u_\mathrm{A}^2 + 2\beta^2 - (1 + (4 - u_\mathrm{A}) u_\mathrm{A})\beta}{(1+u_\mathrm{A})^2} =1+ {\cal O}[\Delta k]^2
\\[0.25cm]
\Lambda_4 &=& -\gamma^2\left(\frac{1-u_\mathrm{A}}{1+u_\mathrm{A}}\frac{1+\beta}{1-\beta}\right)^2(\beta^2-2u_\mathrm{A}\beta +1) = {\cal O}[\Delta k]^2\\[0.25cm]
\Lambda_5 &=& 4\frac{u_\mathrm{A}-\beta}{(1+u_\mathrm{A})^2}\frac{1-u_\mathrm{A}\beta}{(1-\beta)^2} 
 =1- {\cal O}[\Delta k]^2 \\[0.25cm]
\Lambda_6 &=& \left(\frac{1-u_\mathrm{A}}{1+u_\mathrm{A}}\frac{1+\beta}{1-\beta}\right)^2 = {\cal O}[\Delta k]^2 \\[0.25cm]
\Lambda_7 &=& \frac{4u_\mathrm{A}}{(1+u_\mathrm{A})^2}\frac{1+u_\mathrm{A}\beta}{1-\beta} = 1+ {\cal O}[\Delta k] \\[0.25cm]
\Lambda_8 &=& \left(\frac{1+u_\mathrm{A}}{1-u_\mathrm{A}}\right)^2\frac{\gamma^2u_\mathrm{A} (1+\beta)^4}{u_\mathrm{A}-2\beta+u_\mathrm{A}\beta^2} = {\cal O}[\Delta k]^2\\[0.25cm]
\Lambda_{9} &=& 4\gamma^2 \left(\frac{u_\mathrm{A} -\beta}{1+u_\mathrm{A}}\right)^2= 1+{\cal O}[\Delta k] \\[0.25cm]
\lambda&=& \frac{h}{4} \frac{1+u_\mathrm{A}}{1-u_\mathrm{A}}\frac{1-\beta}{1+\beta} = \gamma^2 (1-u_\mathrm{A}\beta) \Lambda\\[0.25cm]
\kappa &=& -\frac{\omega_\mathrm{g}}{u_\mathrm{A}}\frac{u_\mathrm{A} (1+\beta^2) -\beta(1+u_\mathrm{A}^2)}{u_\mathrm{A}(1+\beta^2) -2\beta} 
\simeq \frac{\omega_\mathrm{g}}{u_\mathrm{A}}\left(1+ {\cal O}[\Delta k]\right)\\[0.25cm]  
\end{array}
$$

\bibliographystyle{apj}
\bibliography{aa} 

\end{document}